\documentclass[fleqn,usenatbib]{mnras}

\usepackage{newtxtext,newtxmath}
\usepackage[T1]{fontenc}

\DeclareRobustCommand{\VAN}[3]{#2}
\let\VANthebibliography\thebibliography
\def\thebibliography{\DeclareRobustCommand{\VAN}[3]{##3}\VANthebibliography}

\usepackage{amsmath}
\usepackage{natbib}
\usepackage{graphicx}
\usepackage{tablefootnote}
\usepackage{multirow}
\usepackage{soul}
\usepackage{float}
\usepackage{ulem}
\usepackage[dvipsnames]{xcolor}
\usepackage{multirow}

\newcommand{\detections}{9}
\newcommand{\predicteddetections}{14$\pm$6}
\newcommand{\lsolar}{L$_{\odot}$}
\newcommand{\msolar}{M$_{\odot}$}

\title[Spitzer SN Survey]{A Spitzer Survey for Dust-Obscured Supernovae}
\author[O. D. Fox, et al.]{Ori D. Fox$^{1,2}$, Harish Khandrika$^{1}$, David Rubin$^{3}$, Chadwick Casper$^{4,5}$, \newauthor Gary Z. Li$^{4,6}$, Tam\'as Szalai$^{7,8}$, Lee Armus$^{9}$, Alexei V. Filippenko$^{4,10}$, \newauthor Michael F. Skrutskie$^{11}$, Lou Strolger$^{1}$, Schuyler D. Van Dyk$^{9}$\\
$^{1}$Space Telescope Science Institute, 3700 San Martin Drive, Baltimore, MD 21218, USA.\\
$^{2}$ofox@stsci.edu.\\
$^{3}$Department of Physics and Astronomy, University of Hawai`i at M{\=a}noa, Honolulu, Hawai`i 96822, USA.\\
$^{4}$Department of Astronomy, University of California, Berkeley, CA 94720-3411, USA.\\
$^{5}$Glint Photonics, 1520 Gilbreth Rd., Burlingame, CA 94010, USA.\\
$^{6}$The Aerospace Corporation, 2310 E. El Segundo Blvd., El Segundo, CA 90245, USA.\\
$^{7}$Department of Optics and Quantum Electronics, University of Szeged, H-6720 Szeged, D\'om t\'er 9, Hungary.\\
$^{8}$Konkoly Observatory, Research Centre for Astronomy and Earth Sciences, H-1121 Budapest, Konkoly Thege Mikl\'os \'ut 15-17, Hungary.\\
$^{9}$IPAC, Caltech, 1200 E. California Blvd., Pasadena, CA 91125, USA.\\
$^{10}$Miller Institute for Basic Research in Science, University of California, Berkeley, CA 94720, USA.\\
$^{11}$Department of Astronomy, P.O. Box 3818, University of Virginia, Charlottesville, VA 22903-0818, USA.}

\begin{document}

\maketitle
\begin{abstract} 

Supernova (SN) rates serve as an important probe of star-formation models and initial mass functions. Near-infrared seeing-limited ground-based surveys typically discover a factor of 3--10 fewer SNe than predicted from far-infrared (FIR) luminosities owing to sensitivity limitations arising from both a variable point-spread function (PSF) and high dust extinction in the nuclear regions of star-forming galaxies.  This inconsistency has potential implications for our understanding of star-formation rates and massive-star evolution, particularly at higher redshifts, where star-forming galaxies are more common. To resolve this inconsistency, a successful SN survey in the local universe must be conducted at longer wavelengths and with a space-based telescope, which has a stable PSF to reduce the necessity for any subtraction algorithms and thus residuals. Here we report on a two-year {\it Spitzer}/IRAC 3.6\,$\mu$m survey for dust-extinguished SNe in the nuclear regions of forty luminous infrared galaxies (LIRGs) within 200\,Mpc.  The asymmetric {\it Spitzer} PSF results in worse than expected subtraction residuals when implementing standard template subtraction.  Forward-modeling techniques improve our sensitivity by several $\sim 1.5$ magnitudes.  We report the detection of \detections\ SNe, five of which were not discovered by optical surveys.  After adjusting our predicted rates to account for the sensitivity of our survey, we find that the number of detections is consistent with the models.  While this search is nonetheless hampered by a difficult-to-model PSF and the relatively poor resolution of {\it Spitzer}, it will benefit from future missions, such as {\it Roman} and {\it JWST}, with higher resolution and more symmetric PSFs.

\end{abstract}


\section{Introduction}
\label{intro}

The observed rate at which stars more massive than $\sim 8$ \msolar\ explode as core-collapse supernovae (CCSNe) can be used to determine chemical evolution and feedback processes \citep{matteucci06, scannapieco05}, progenitor-mass distributions \citep{smith11}, star-formation rates \citep{iben83, mannucci07a}, and dust yields \citep{maiolino04a, maiolino04b, rho09}.  Given the intrinsic brightness of SNe, they serve as useful probes of the above characteristics at higher redshifts where other techniques are less feasible \citep[e.g.,][]{dahlen99,dahlen04,strolger15}.

SN rates are useful probes, however, only if we understand the models linking the initial mass function (IMF), star-formation rates (SFRs), and SNe.  For example, \citet{mattila01} derive the expected CCSN rate (CCSNr) as a function of a galaxy's far-infrared (FIR) luminosity, $L_{\rm FIR} = L_{{\rm (8-1000)}~\micron}$ (which is used as a proxy for star formation):
\begin{equation}
{\rm CCSNr} = \frac{2.7\,L_{\rm FIR}}{10^{10}~{\rm L_{\odot}}} {\rm SN}\,(100~{\rm yr})^{-1}.\\
\label{eqn_rate}
\end{equation}

\begin{figure}
\centering
\includegraphics[height=6.2cm]{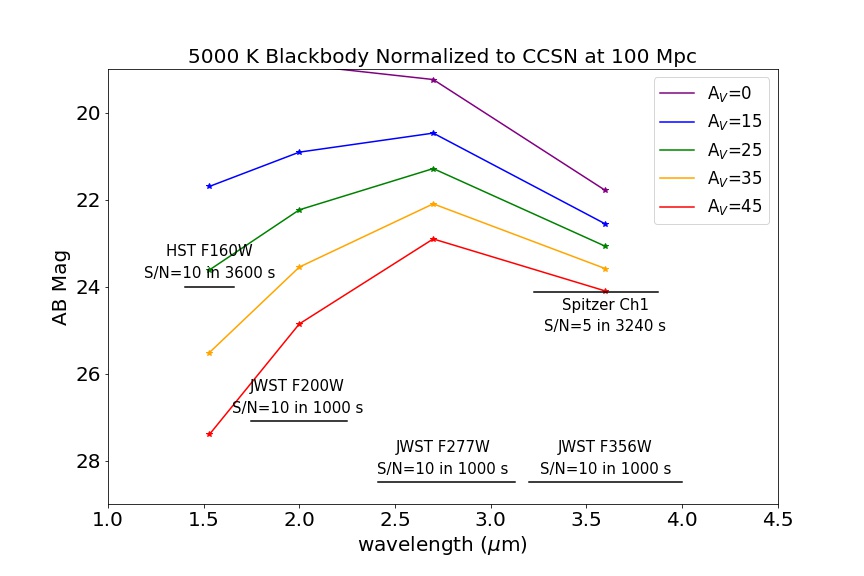}
\caption{The effects of dust on the intensity and shape of a blackbody spectrum.  The SED here is that of a 5000\,K blackbody assumed to represent the photosphere of a SN~IIP several months post-explosion.  The 3.6\,\micron\ flux is normalised to $M=-16$\,mag Vega (see Fig.~\ref{fig_lumfunction}).  Since more absorption occurs at shorter wavelengths, the peak of the spectrum shifts beyond the NIR (i.e., 2.6\,\micron) as $A_V$ increases beyond 25\,mag.  {\it Spitzer} is more efficient than {\it HST} at these higher extinction levels.}
\label{fig_extinction}
\end{figure}


\noindent While Equation \ref{eqn_rate} is an empirical relationship, \citet{mattila01} derive a similar relation (their Eq.~5) from first principles and show that Equation \ref{eqn_rate} both has a basis in and is consistent with the theoretical connection between the FIR luminosity and SFR.  \citet{mattila12} add that this empirical relationship could have a significant uncertainty owing to a small sample size.  We further note that the theoretically derived relationship also has significant uncertainty since it assumes a scaling factor between the intrinsic FIR-specific luminosity and the SFR, with no correction for extinction or other effects\footnote{This includes the an unknown efficiency in reprocessing UV radiation from stars into dust emission in the far-IR, which for starbursts with high optical depths is around $100\%$.}, of about $\kappa_{\rm FIR}\approx4\times10^{-10}\, {\rm M}_{\odot}\, {\rm yr}^{-1}\, {\rm L}_{\odot}^{-1}$. This is about twice as large as the value determined for dusty galaxies by \citet{madau14} and \citet{kennicutt98}, at ages of above around 300 Myr.\footnote{Younger galaxies ($\lesssim 300$ Myr) have higher emission-to-star-formation ratios as short-lived massive stars still dominate UV emission that is reprocessed and re-emitted in the IR.} The derivation also assumes the fraction of stars ultimately successful in producing CCSNe is $k \approx 0.007~{\rm M}_{\odot}^{-1}$.  This value is $\sim30$\% larger than estimated for average galaxies by \citet{strolger15}, but also carries some assumptions on progenitor mass ranges an/d average shape of the high-mass IMF.  Considering the relationship between all of these variables, the observed CCSNr therefore gives a useful constraint on the model by providing some leverage on our theoretical understanding of $k$ and $\kappa_{\rm FIR}$.\\

The ability to detect all CCSNe in a survey limits the completeness of any rate study.  This is particularly true in galaxies with high SFRs, where gas and dust obscure many CCSNe.  In fact, optical-wavelength surveys in dusty, star-forming (ultra)-luminous IR galaxies (LIRGs and ULIRGs) have routinely established CCSNr values that are unexpectedly similar to those in more normal galaxies \citep{richmond98, navasardyan01}.  \citet{strolger15} derive the CCSNr from the {\it Hubble Space Telescope (HST)} CANDELS and CLASH programs, two extragalactic multicycle treasury programs using the WFC3 IR channel with sensitivities down to 25.5 mag$_{\rm Vega}$ out to 1.6 \micron.  The observations are notably different from the ultraviolet (UV)/optical SFR \citep{madau14}, and even less so with the mid-IR (MIR) and FIR predictions \citep{chary01}.  \citet{strolger15} link the low CCSNr (and the subsequent source of discrepancy between the UV/optical and IR-derived SFRs) to dust obscuration, particularly in galaxies with high SFRs, where gas and dust obscure most of the SNe, in agreement with \citet{mattila12}.

The highest SFRs in the local Universe are found in LIRGs and ULIRGs ($L_{\rm FIR}>10^{11}$~\lsolar\ and $10^{12}$~\lsolar, respectively; \citealt{sanders03}).  Although one must account for the contribution from hidden active galactic nuclei (AGNs), star formation tends to contribute most of the IR luminosity \citep{sanders96, genzel98}, and estimates of the relative AGN power have been made using a variety of MIR tracers in hundreds of local IR galaxies \citep[e.g.,][]{diaz-santos17}.  (U)LIRGs account for only a small fraction of the local ($<100$ Mpc) galaxy population \citep{sanders03}, so obscuration in these galaxies does not have a detrimental effect on {\it local} CCSNr studies.  At higher redshifts, however, obscured star formation in (U)LIRGs {\it dominates} over star formation traced by UV and optical light \citep[e.g.,][]{floch05,magnelli09,magnelli11,strolger15}.  To accurately measure the CCSNr as a function of redshift, the local CCSNr in starburst galaxies, such as (U)LIRGs, must be fully characterised.

\begin{figure}
\centering
\vbox{
    \raisebox{0.0in}{\includegraphics[width=3.3in]{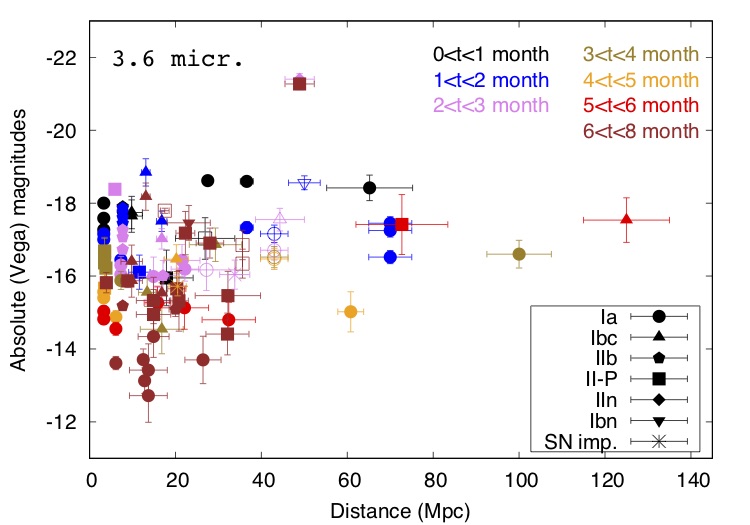}}
    }
\caption{Early-time MIR evolution of all historical SNe obtained from {\it Spitzer} archival data by \citet{szalai19}.  Filled and open symbols denote SNe where photometry was obtained with or without image subtraction, respectively. SN subtypes \citep[e.g.,][]{filippenko97} are denoted in the legend, and ``SN imp" refers to SN impostors.}
\label{fig_lumfunction}
\end{figure}

\begin{table*}
\centering
\caption{Target Galaxies}
\begin{tabular}{lllrrrrr}
Name &            RA (J2000) &           Dec (J2000) &  Distance &  $L_{\rm FIR}$ &   \# of SNe &  \# of SNe &  Detections \\
    &   &   &  (Mpc) &  (log \lsolar) & & (Corrected) & \\
\hline
NGC 34           &  00:11:06.612 &  -12:06:28.33 &     83.33 & 11.43 &  1.45 &           0.42 &           0 \\
NGC 232          &  00:42:45.814 &  -23:33:40.69 &     91.66 & 11.23 &  0.92 &           0.28 &           0 \\
MCG+12-02-001    &  00:54:03.943 &  +73:05:05.23 &     66.67 & 11.29 &  1.05 &           0.71 &           0 \\
IC 1623          &    01:07:47.2 &     -17:30:25 &     79.17 & 11.38 &  1.30 &           0.35 &           0 \\
UGC 2369         &    02:54:00.9 &     +14:58:31 &    129.17 & 11.44 &  1.49 &           0.09 &           0 \\
IRAS 03359+1523  &    03:38:46.9 &     +15:32:55 &    145.83 & 11.38 &  1.30 &           0.04 &           1 \\
MCG-03-12-002    &    04:21:20.0 &     -18:48:45 &    133.33 & 11.30 &  1.08 &           0.05 &           0 \\
NGC 1572         &  04:22:42.814 &  -40:36:03.50 &     83.33 & 11.16 &  0.78 &           0.34 &           0 \\
NGC 1614         &  04:34:00.027 &  -08:34:44.57 &     66.67 & 11.41 &  1.39 &           0.56 &           0 \\
NGC 2623         &  08:38:24.093 &  +25:45:16.70 &     75.00 & 11.47 &  1.59 &           0.82 &           0 \\
UGC 4881         &   09:15:55.54 &   +44:19:58.2 &    166.67 & 11.57 &  2.01 &           0.01 &           0 \\
UGC 5101         &  09:35:51.694 &  +61:21:10.52 &    162.50 & 11.90 &  4.29 &           0.02 &           0 \\
MCG+08-18-012    &    09:36:30.7 &     +48.28:10 &    108.33 & 11.19 &  0.84 &           0.14 &           0 \\
IC 563/IC 564    &  09:46:20.361 &  +03:02:43.86 &     83.33 & 11.10 &  0.68 &           0.29 &           1 \\
NGC 3110         &  10:04:02.124 &  -06:28:29.12 &     66.67 & 11.10 &  0.68 &           0.44 &           0 \\
NGC 3256         &   10:27:51.60 &   -43:54:18.0 &     37.50 & 11.44 &  1.49 &           1.25 &           1 \\
IRAS 10565+2448  &  10:59:18.153 &  +24:32:34.30 &    175.00 & 11.87 &  4.00 &           0.03 &           0 \\
Arp 148          &  11:03:53.892 &  +40:50:59.89 &    141.67 & 11.50 &  1.71 &           0.06 &           1 \\
MCG+00-29-0023   &  11:21:12.261 &  -02:59:03.00 &    100.00 & 11.36 &  1.24 &           0.26 &           0 \\
IC 2810/UGC 6436 &  11:25:45.055 &  +14:40:35.98 &    141.67 & 11.50 &  1.71 &           0.06 &           0 \\
NGC 3690         &   11:28:33.13 &   +58:33:58.0 &     45.83 & 11.72 &  2.83 &           0.24 &           0 \\
ESO507-G070      &  13:02:52.354 &  -23:55:17.65 &     87.50 & 11.31 &  1.10 &           0.39 &           0 \\
UGC 8335         &    13:15:32.8 &     +62:07:37 &    129.17 & 11.60 &  2.15 &           0.06 &           0 \\
UGC 8387         &  13:20:35.380 &  +34:08:21.84 &     95.83 & 11.52 &  1.79 &           0.23 &           0 \\
NGC 5256         &   13:38:17.69 &   +48:16:33.9 &    112.50 & 11.37 &  1.27 &           0.16 &           0 \\
NGC 5257         &  13:39:52.273 &  +00:50:22.48 &     91.67 & 11.37 &  1.27 &           0.40 &           1 \\
Mk 273           &  13:44:42.070 &  +55:53:13.17 &    158.33 & 12.10 &  6.80 &           0.08 &           0 \\
NGC 5331         &   13:52:16.15 &   +02:06:03.3 &    137.50 & 11.43 &  1.45 &           0.06 &           0 \\
UGC 8782         &    13:52:17.7 &     +31:26:44 &    187.50 & 12.27 & 10.06 &           0.00 &           0 \\
Arp 302          &    14:57:00.4 &     +24:36:44 &    141.67 & 11.59 &  2.10 &           0.05 &           0 \\
Mk 848           &  15:18:06.123 &  +42:44:44.59 &    166.67 & 11.72 &  2.83 &           0.03 &           0 \\
Arp 220          &  15:34:57.272 &  +23:30:10.48 &     75.00 & 12.12 &  7.12 &           3.52 &           0 \\
NGC 6090         &   16:11:40.39 &   +52:27:21.5 &    120.83 & 11.35 &  1.21 &           0.11 &           1 \\
NGC 6240         &   16:52:58.97 &   +02:24:01.7 &    100.00 & 11.85 &  3.82 &           0.65 &           2 \\
IRAS 17208-0014  &  17:23:21.943 &  -00:17:00.96 &    179.17 & 12.30 & 10.77 &           0.05 &           0 \\
IC 4687/86       &  18:13:39.829 &  -57:43:31.25 &     70.83 & 11.35 &  1.21 &           0.68 &           0 \\
IRAS 18293-3413  &    18:32:40.2 &     -34:11:26 &     75.00 & 11.63 &  2.30 &           0.97 &           1 \\
NGC 6926         &  20:33:06.108 &  -02:01:39.07 &     83.33 & 11.11 &  0.70 &           0.29 &           0 \\
NGC 7130         &  21:48:19.490 &  -34:57:04.73 &     66.67 & 11.21 &  0.88 &           0.48 &           0 \\
IRAS 23128-5919  &  23:15:46.772 &  -59:03:15.94 &    183.33 & 11.80 &  3.41 &           0.01 &           0 \\
\end{tabular}

\label{tab1}
\end{table*}

\begin{figure*}
\centering
\includegraphics[width=6.0in]{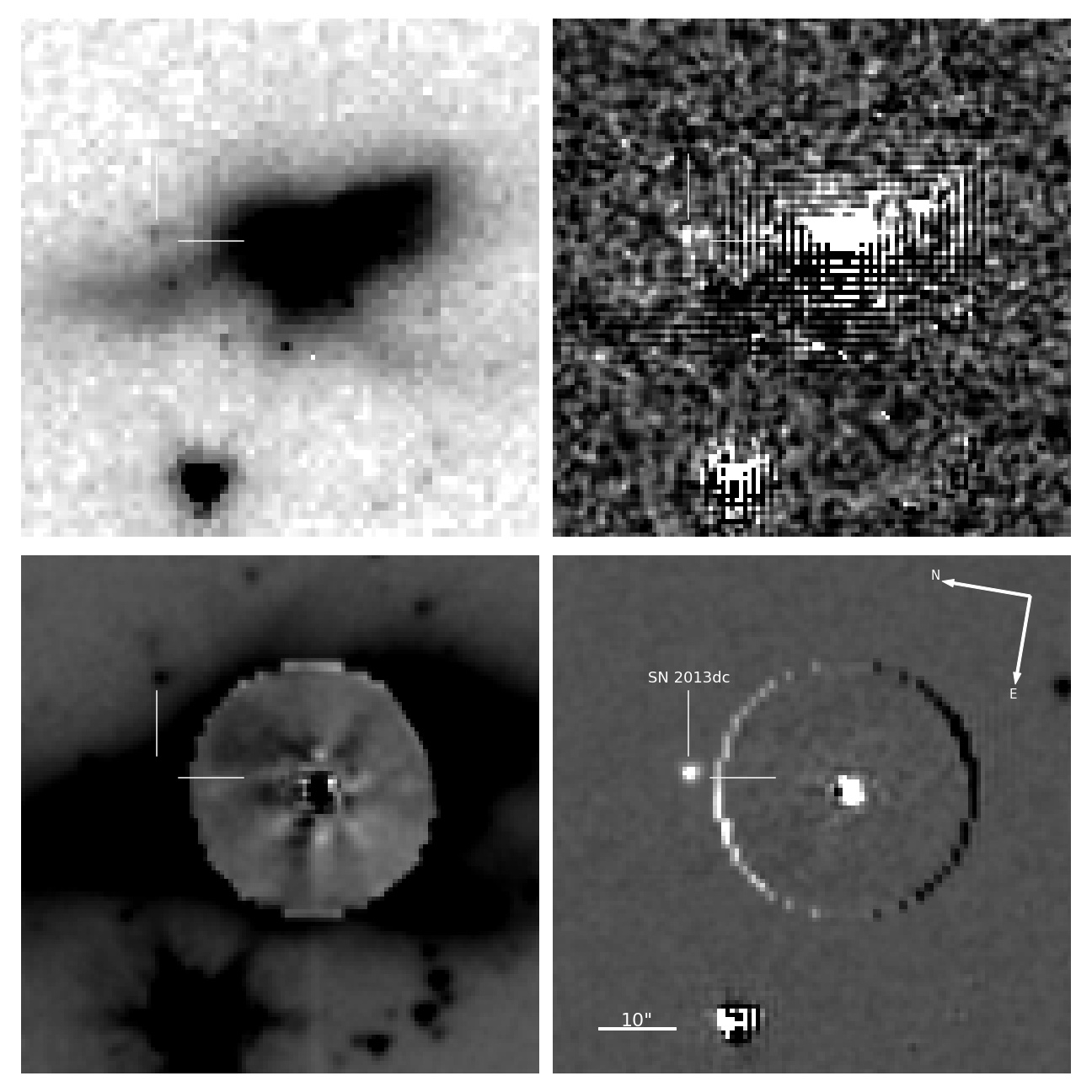}
\caption{Comparison of sample data products, in this case for the first epoch of NGC 6240, used in our transient search. ({\it Top left:}) Original, unsubtracted MOPEX processed data. ({\it Top right:}) Standard alignment and subtraction of two epochs. ({\it Bottom left:}) Forward modeling subtraction, where the circle corresponds to our forward modeling radius. ({\it Bottom right:}) Subtraction of two epochs that have been forward modeled, but in this case are spaced nearly one year apart so that the telescope roll angle is similar.  The colour scale and stretch is constant for all subtracted images shown in this figure. The point source is SN 2013dc.  It is present in all the images, but is most clearly detected in the bottom right.  The central source is the galaxy nucleus.  As described in the text, there are a number of caveats for modeling the inner core, including nonlinearity concerns.  Correcting those issues is beyond the scope of this paper, but we account for them in our statistics.}
\label{fig_forward_modeling} 
\end{figure*}

The IR, which can be up to 10 times less affected by extinction than visible light, offers an optimised window for SN searches in these starburst galaxies.  Already, a number of near-IR (NIR) searches have been performed with mixed success.  Early NIR ground-based surveys were hampered by poor resolution and limited telescope time \citep{buren94,grossan99,maiolino02,mannucci03,mattila04, mattila05a,mattila05b,miluzio13}.  These studies resulted in an inferred CCSNr that was still a factor of 3--10 lower than expected from the SFR \citep{mannucci03}.  Furthermore, no SNe were detected within the galaxy nuclei.  Ground-based high-resolution adaptive optics (AO) NIR searches have had some additional success in their discovery of several CCSNe within $<300$\,pc from the LIRG nuclei \citep[e.g.][]{mattila07, kankare08, kankare12, kool18, kankare21,perez21}. \citet{kool18} note their high-spatial-resolution search uncovered a larger concentration of nuclear SNe in (U)LIRGs than previous low-resolution searches, while \citet{kankare21} showed that a sample of 29 CCSNe at $<$2.5 kpc from (U)LIRG nuclei showed a correlation between the startburst age and SN subtype.  In general, these gorund-based surveys either did not find their rates consistent with the predictions or had limited statistics.   The overall conclusion was that a majority of SNe likely occur within the nuclei, but the extinction is so high ($A_V > 25$\,mag) and the nuclei are so bright that longer-wavelength observations are necessary.  Even several space-based surveys \citep[e.g.,][]{bregman00}, including a {\it HST}/NICMOS \citep{cresci07}, turned up no confirmed detections, concluding that the ULIRG dust-extinction values are too high ($A_V > 25$\,mag).  Results from the SPitzer InfraRed Intensive Transients Survey (SPIRITS; \citealt{kasliwal17}) between 2014 and 2018 find the fraction of CCSNe in nearby galaxies missed by optical surveys could be as high as 38.5\% \citep{jencson19}. High-resolution ground-based radio surveys, by contrast, have successfully discovered nuclear SNe with rates consistent with the galaxy IR luminosities, but not all SNe are sufficiently bright at radio wavelengths to conduct a complete survey \citep{perez21}.

Until the {\it James Webb Space Telescope (JWST)} is launched, the {\it Spitzer Space Telescope (Spitzer)} Infrared Array Camera (IRAC; \citealt{fazio04}) offered the best combination of long-wavelength sensitivity, stable point-spread function (PSF), moderate resolution, and sensitivity to SN radiation.  Here we describe a post-cryogenic Warm-{\it Spitzer}/IRAC survey conducted in the years 2012--2014 for dust-extinguished SNe in, but not limited to, the nuclear regions of nearby star-forming (U)LIRGs (PID 90031; PI: O.~Fox). (A similar, but more limited survey was conducted during the cryogenic {\it Spitzer\/} mission by \citealt{lawrence04}, PID 108; however, the data analysis and results were never published.) The MIR is optimised for dust-extinction levels $A_V > 25$\,mag.  The improved sensitivity offers an improved estimate of the number of SNe missed by visible (and NIR) surveys and tighter constraints on our understanding of star-formation models out to high redshifts.  The direct product of this study is the connection between FIR luminosity and massive-star formation.  Such results naturally connect to the high-redshift universe, where a fundamental observable is the FIR galactic flux. Given a future which includes the LSST, {\it EUCLID}, the {\it Nancy Grace Roman Space Telescope (NGRST)}, and {\it JWST}, local measurements characterising the properties of galaxies selected from FIR samples have large-scale implications. 

Our observations are presented in  Section~\ref{sec_obs}. Section~\ref{sec_reduction} describes the data reduction, data processing, and analysis techniques. Our results are discussed in Section~\ref{sec_results}, and Section~\ref{sec_conclusions} summarises our conclusions.

\section{Observations}\label{sec_obs}

The original proposal called for a Warm-{\it Spitzer}/IRAC survey of 40 galaxies to maximise the number of expected SNe and generate a statistically significant sample to minimise the statistical effects of undetected SNe.  We selected the nearest 40 (U)LIRGs ($L_{\rm FIR} > 10^{10}$~\lsolar) from \citet{sanders03}, which ultimately extended our sample out to 200 Mpc. Table~\ref{tab1} summarises the targets and also lists the number of SNe predicted by Equation~\ref{eqn_rate} over the two-year survey window.  For each galaxy, eight epochs of observations were obtained, spaced approximately 1--6 months apart depending on the visibility window.

To minimise observing time, all observations were obtained in only a single filter with IRAC Channel 1 (3.6\,\micron).  The spectral energy distribution (SED) for a 5000\,K blackbody peaks at $>2.0$\,\micron\ for $A_V > 25$\,mag (Fig. \ref{fig_extinction}).  Increasing the extinction pushes the peak of the SED to longer wavelengths, but also lowers the overall flux substantially enough that it becomes inefficient to observe these targets at larger distances.  Channel 1 (3.6\,\micron) is also slightly more sensitive than Channel 2 (4.5\,\micron).  These are the only two channels available during the Warm-{\it Spitzer} mission and, at the time of the survey, {\it Spitzer}/IRAC was the only MIR observing capabilities accessible from space.

\begin{figure}
\centering
\includegraphics[width=3.in]{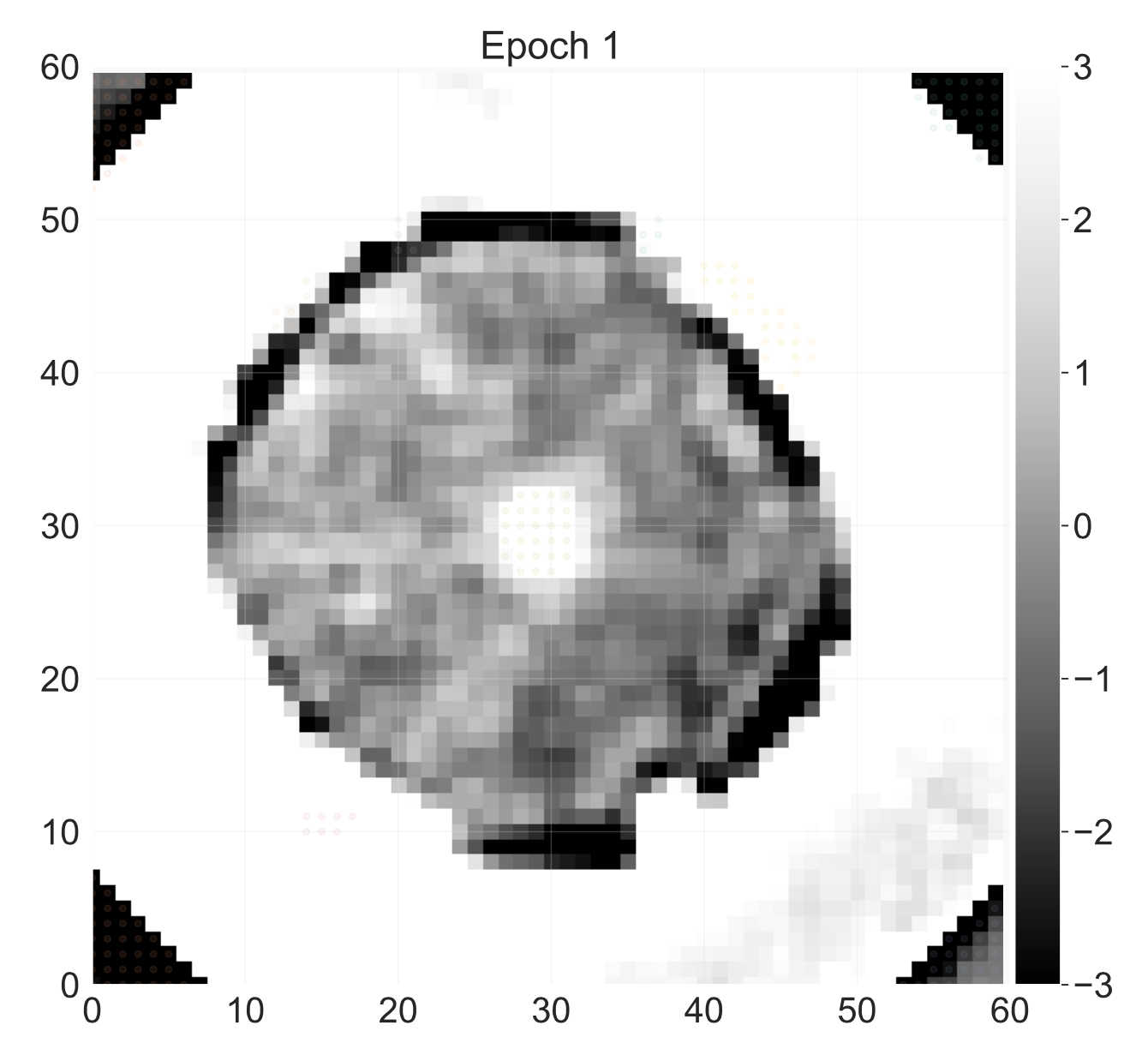}
\caption{A residual map of UGC 8782 illustrating complications and limitations of our algorithm to detect interconnected groups of pixels within a given epoch with fluxes $\gtrsim3\sigma$ of their average value.  The original image has the forward model subtracted, resulting in the ringed structure corresponding to the model itself.  The color scale units are $\sigma$, defined by the statistics of each individual pixel across all eight epochs.  The poorly subtracted nuclear region in this example is typical of many of the targets in our sample for a variety of issues beyond forward modeling that we describe and account for in the text.}
\label{fig_threesig} 
\end{figure}

To calculate our integration times, we consider the distribution of CCSN magnitudes.  MIR light curves of CCSNe are sparse, but \citet{szalai19} present comprehensive plots of all existing {\it Spitzer} observations.  Figures 6 and 7 from \citet{szalai19} show that CCSNe plateau in the MIR for $\sim 300$ days post-explosion, and often even longer.  Figure~\ref{fig_lumfunction} plots the photometry as a function of distance, separated by explosion type and age.  A majority of CCSNe fall within a range of $-16 < M_{3.6\,\mu m} < -18$ over the first few months.  Most of these SNe are at distances $<60$ Mpc, since the proposed {\it Spitzer} observations generally targeted nearby, bright SNe.

\begin{figure}
\centering
\includegraphics[width=3.5in]{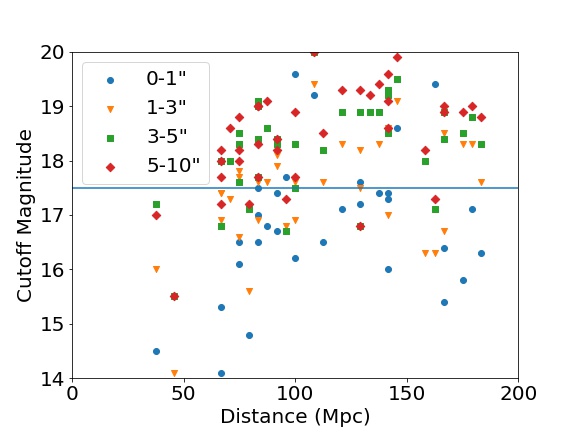}
\caption{Sensitivity of our search algorithm derived from the artificial source tests for each galaxy as a function of the galaxy distance.  Each galaxy in our sample has four data points, corresponding to recovery rates for artificial sources placed at various radii from the nucleus as defined by the legend.  The detection limit, or cutoff magnitude, is defined as the magnitude at which we can achieve a 50\% recovery rate for all of our sources.  The solid line corresponds to the more realistic empirical cutoff magnitude derived qualitatively from Figure \ref{fig_detection_mags}, as we discuss in the text.  Figure \ref{fig_overallpolygon2} plots the consolidated results for all galaxies shown in this plot.}
\label{fig_cutoffmag} 
\end{figure}

Given that the typical time between consecutive observations is anywhere in the range 1--6 months, we assumed a magnitude of $M_{3.6\,\mu m} \approx -17$ when designing our observations.  In general, photon-counting statistics from the underlying galaxy dominate the uncertainty, and we therefore require longer integration times. We also account for possible extinction ranging up to $A_V=35$\,mag.  Figure~\ref{fig_extinction} shows the effects of extinction by plotting the SED for a 5000\,K blackbody for various values of $A_V$.  At 3.6\,\micron, the effects of extinction are minimal, decreasing the brightness only $\sim 2$\,mag for $A_V=45$\,mag.  Despite this planning, however, subtraction effects ultimately dominated our error budget, which we discuss in more detail below.

\section{Reduction Techniques}\label{sec_reduction}

\subsection{Forward Modeling and Subtraction}

Despite the absence of atmospheric effects when working with images taken by space telescopes, there are three features of the {\it Spitzer} data that combine to make precise galaxy subtraction difficult. (1) The PSF is undersampled (only $\sim 1$ pixel per full width at half-maximum intensity). (2) The PSF is highly azimuthally asymmetric. (3) The data are taken over a range of spacecraft roll angles, with no two epochs matching exactly. To enable  precise galaxy subtraction, we turn to forward modeling.

Forward modeling infers an analytic galaxy model on the sky by tracing the light (``forward'') through convolution with the PSF and pixel, and then sampling by the detector \citep[e.g.,][]{holtzman08,suzuki12,hayden21,Rubin2021}. Each of the epochs is fit simultaneously with a model of the galaxy background plus SN.  This model is inferred using a set of dithers and/or rotations, bypassing the traditional step of resampling the images for pixel-by-pixel subtraction, which would introduce artifacts. In testing, we found that the given model of the Point Response Function\footnote{https://irsa.ipac.caltech.edu/data/SPITZER/docs/irac/calibrationfiles/psfprf/} (PRF, the covolution of the PSF and the pixel response) was not accurate enough for our purposes, so we derived our own PRF from high signal-to-noise-ratio observations of field stars.

For each of the eight epochs in which a search is performed, the forward model is derived from reference observations of the galaxy obtained before the SN appears or after the SN fades. In other words, we exclude observations obtained $< 20$ days prior or $> 330$ days to account for the possibility that a rising/falling SN component may contribute to the model.  In addition to the galaxy model, we also fit small tweaks to the astrometry and differences in the sky level.  Once a model is generated, it is subtracted from each corrected Basic Calibrated Data (cBCD) file, which are then stacked with the MOsaicker and Point source EXtractor (MOPEX)\footnote{https://irsa.ipac.caltech.edu/data/SPITZER/docs/dataanalysistools/tools/mopex/} (Fig. \ref{fig_forward_modeling}). 

\begin{figure}
\centering
\includegraphics[width=3.5in]{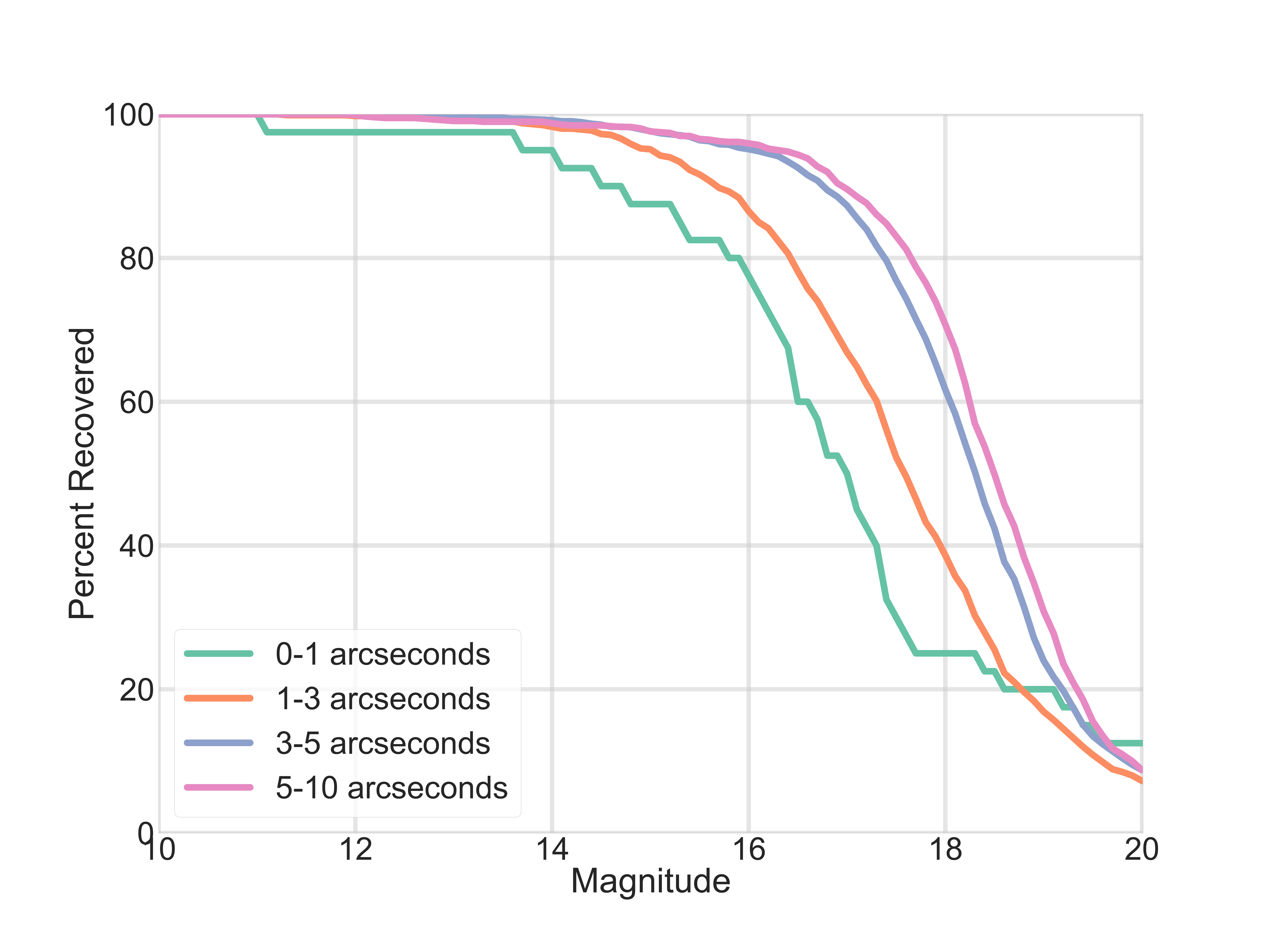}
\caption{Overall detection rate as a function of radius from the galactic nucleus for forward model subtracted images based on artificial star tests}.
\label{fig_overallpolygon2} 
\end{figure}

\subsection{Detecting Transient Sources}
\label{sec_algorithm}

To quantify the success of the forward-modeling algorithm, we use grids of artificial star tests.  Owing to the IRAC mosaic's large pixel scale ($\sim0.6$\arcsec), however, the flux density within any single pixel tends to vary from epoch to epoch, particularly in rapidly changing (i.e., nuclear) environments.  This could lead to  false positives.  We therefore developed an approach that considers the variance in each pixel over all eight epochs and then searches for a cluster of interconnected pixels within a given epoch that are all $>3\sigma$ from their average value (e.g., Fig. \ref{fig_threesig}).  For each galaxy, we calculate the percent of SNe recovered using this method as a function of magnitude and distance from the centre of the galaxy.  The detection limit, or cutoff magnitude, is defined as the magnitude at which we can achieve a 50\% recovery rate and varies significantly depending on the individual features of the host galaxy (Fig. \ref{fig_cutoffmag}).

\begin{figure}
\centering
\includegraphics[width=3.5in]{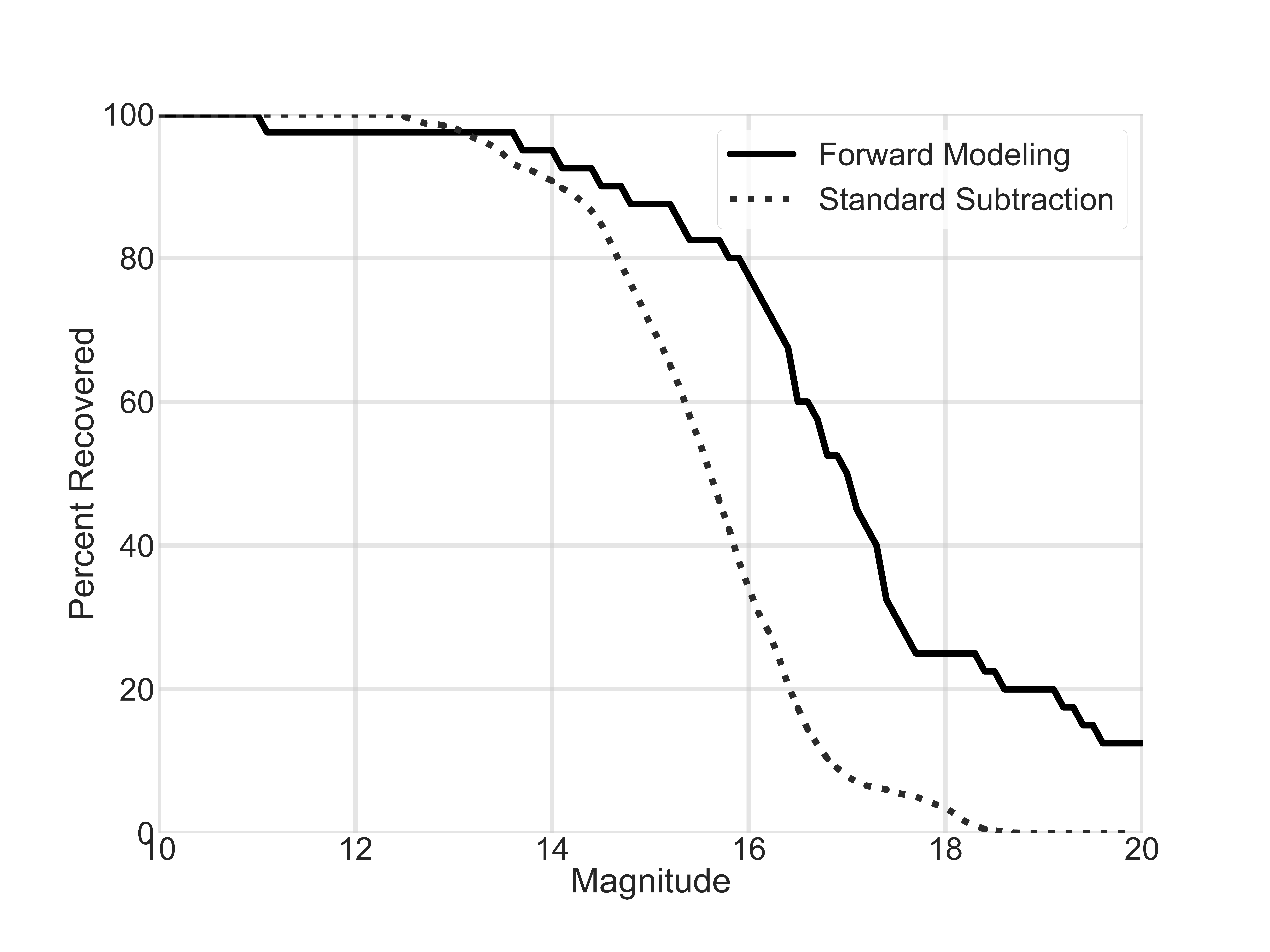}
\caption{Overall detection rate for 0--5\arcsec\ comparing the standard and forward model subtraction techniques based on artificial star tests}.
\label{fig_comparison}
\end{figure}

Figures~\ref{fig_overallpolygon2} and \ref{fig_comparison} consolidate the results from Figure \ref{fig_cutoffmag} and show the evolution of the detection rate as a function of magnitude for the  different radii bins.  Generally, the detection limit improves at larger radii owing to less complicated subtraction residuals.  For comparison, results for the traditional (i.e., non-forward modeling) subtraction images are also plotted in Figure~\ref{fig_comparison}.  Note how the forward modeling improves our relative fraction of faint SNe ($> 14$ mag) recovered. 

Using this $3\sigma$ interconnected pixel detection algorithm, however, is limited given only eight epochs of data per pixel and the fact that SNe can span multiple epochs, thereby influencing the statistics of the pixels that we are trying to isolate.  Any single threshold does not equally apply to all pixels because the asymmetric PSF impacts the noise in the model subtractions of some pixels in a non-Gaussian way. We also notice a subtle nonlinearity in the pixel response that evolves over the two-year lifetime of the survey (i.e., later epochs have lower fluxes).  Pixels may therefore exhibit $3\sigma$ deviations that are not due to Poisson statistics alone (Fig. \ref{fig_threesig}).  Furthermore, we have prior knowledge of the artificial source locations, so recovering the sources is not hampered with false positives or nearby noise effects.  

\begin{figure}
\centering
\includegraphics[width=3.5in]{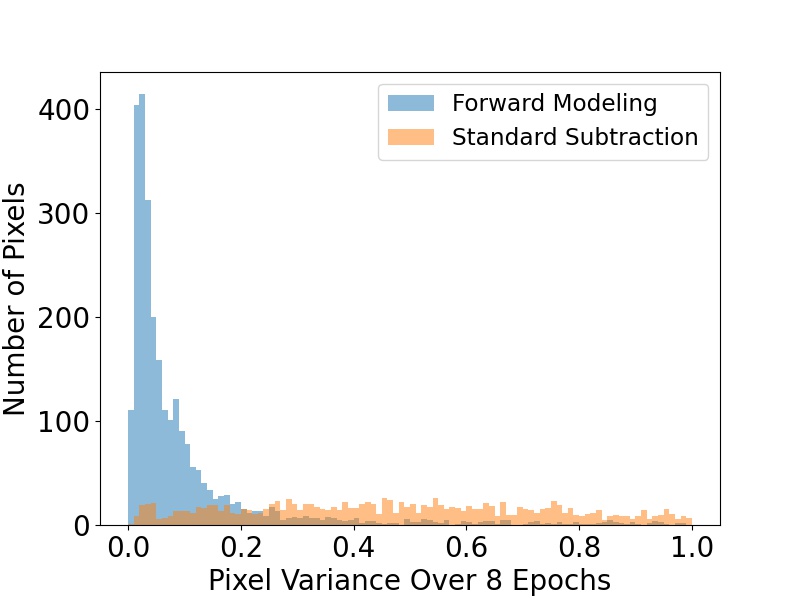}
\caption{Histograms of flux-density variance per pixel in the inner $<$3\arcsec\ of each galaxy using the two different reduction techniques: standard template subtraction and forward modeling.  The histogram for the forward modeling is much narrower, suggesting less variance per pixel and overall increased sensitivity to detecting fainter sources.}
\label{standard_deviation} 
\end{figure}

\begin{table*}
\centering
\caption{Supernova Detections in this Survey}
\begin{tabular}{llccccl}
Host &  Source RA/Dec & Nuclear Offset &  Date & $F_{\nu,[3.6\,\mu m]}$ & $m/M$  &  Notes \\
Galaxy &  (J2000) & (\arcsec N/E) & (MJD) &($\mu$Jy) & ($3.6\,\mu m$)  & \\
\hline
NGC 5257 &  13:39:52.2148, +0:50:25.328 & -11.8/0.3 &  56780 & 12(5) & 18.3/$-$16.5 (0.5) & --\\
\hline
\multirow{4}{*}{IRAS 18293-3413} & \multirow{4}{*}{18:32:41.0980, $-$34:11:27.230} & \multirow{4}{*}{-1.0/0.2} & 56455 & 55(5) & 16.8/$-$17.6 (0.1) & \multirow{4}{*}{SN  2013if} \\
 & & & 56483  & 26(5) & 17.6/$-$16.8 (0.3) & \\
 & & & 56613  & 87(5) & 16.3/$-$18.1 (0.1) & \\
 & & & 56646  & 45(5) & 17.0/$-$17.4 (0.2) & \\
\hline
Arp 148 & 11:03:53.1180, +40:51:09.385 & -8.8/9.5 & 56510 & 1(0) & 21.2/$-$14.6 (0.1) & --\\
\hline
NGC 6090 & 16:11:40.3172, +52:27:23.609 & -3.7/-3.3 & 56890 & 44(5) & 17.0/$-$18.4 (0.2) & -- \\
\hline
\multirow{2}{*}{NGC 3256}  & \multirow{2}{*}{10:27:50.7987, $-$43:54:02.554} & \multirow{2}{*}{-5.2/11.0} & 56875 & 610(10) & 14.1/$-$18.7 (0.1) & \multirow{2}{*}{PSN J10275082-4354034}\\ 
  &  &  & 56925 & 142(20) & 15.7/$-$17.1 (0.2) & \\ 
\hline
IRAS 03359 & 03:38:47.3486, +15:32:56.462 & 3.0/2.7 & 56964 & 24(2) & 17.7/$-$18.1 (0.1) & -- \\
\hline
IC 563 & 09:46:20.7408, +3:02:38.346 & 5.7/-5.5 & 56687 & 31(3) & 17.4/$-$17.2 (0.3) &  -- \\
\hline
\multirow{3}{*}{NGC 6240} & \multirow{3}{*}{16:52:57.6161, +02:23:36.731} & \multirow{3}{*}{-18.7/-26.8} & 56795 & 46(4) & 17.0/$-$18.0 (0.3) & \multirow{3}{*}{PS1-14xw}\\
 & &  & 56833 & 20(5) & 17.8/$-$17.2 (0.4) & \\
 & &  & 56953 & 5(2) & 19.4/$-$15.6 (0.8) & \\
\hline
\multirow{4}{*}{NGC 6240} & \multirow{4}{*}{16:52:58.9791, +02:24:25.587}& \multirow{4}{*}{1.8/22.0} & 56421 & 56(5) & 16.8/$-$18.2 (0.1) & \multirow{4}{*}{SN 2013dc}\\
 & &  & 56454 & 55(5) & 16.8/$-$18.2 (0.1) & \\
 & &  & 56580 & 8(2) & 18.9/$-$16.1 (0.2) & \\
 & &  & 56617 & 7(2) & 19.0/$-$16.0 (0.2) & \\
\hline
\end{tabular}

\label{tab_detections}
\end{table*}

All of this is to say that false positives could impact our interpretation of Figures~\ref{fig_overallpolygon2} and \ref{fig_comparison}.  To alleviate any of these concerns, Figure \ref{standard_deviation} plots the distribution of variance in each pixel for the two different subtraction techniques.  The forward-modeling approach has a much narrower distribution, which is consistent with an improved subtraction approach.  We conclude that the artificial star tests are most useful for guiding the relative success of our forward modeling and subtraction algorithms, but the limitations of the simulations are significant enough that we cannot use them as an absolute quantification of the sensitivity of our search.  Ultimately, we choose to manually search for transient sources in the resulting images with limited guidance from our algorithm. 

\begin{figure*}
\includegraphics[trim=0 0 0 70, clip,height=8.8in]{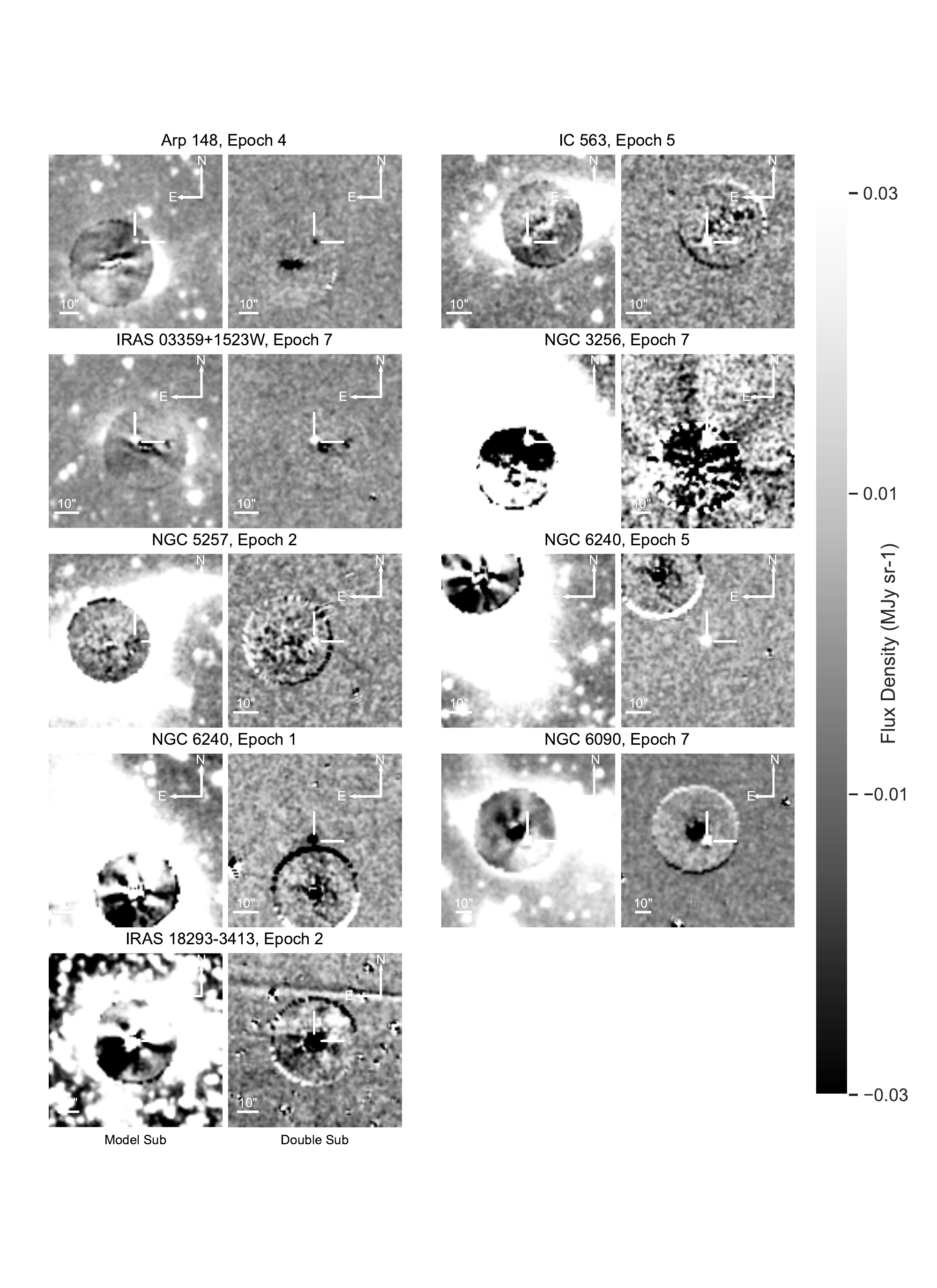}
\vspace{-1.1in}
\caption{Imaging of each SN detected in this survey, labeled by galaxy and epoch.  For each detection, two images are displayed.  On the left is the subtraction of the template derived using forward modeling.  On the right is the subtraction of two different epochs that have been forward modeled.  Epochs separated by $\sim 1$\,yr have similar orientations and the relative subtractions remove additional residuals that were still visible after forward modeling, as well as the more diffuse, outer galaxy that was not forward modeled.  In general, this relative subtraction of epochs separated by about a year provided optimal results.}
\label{fig_detection_images}
\end{figure*}

\section{Results}\label{sec_results}

\subsection{Detections}

\begin{figure}
\centering
\vbox{
    \raisebox{0.0in}{\includegraphics[width=3.5in]{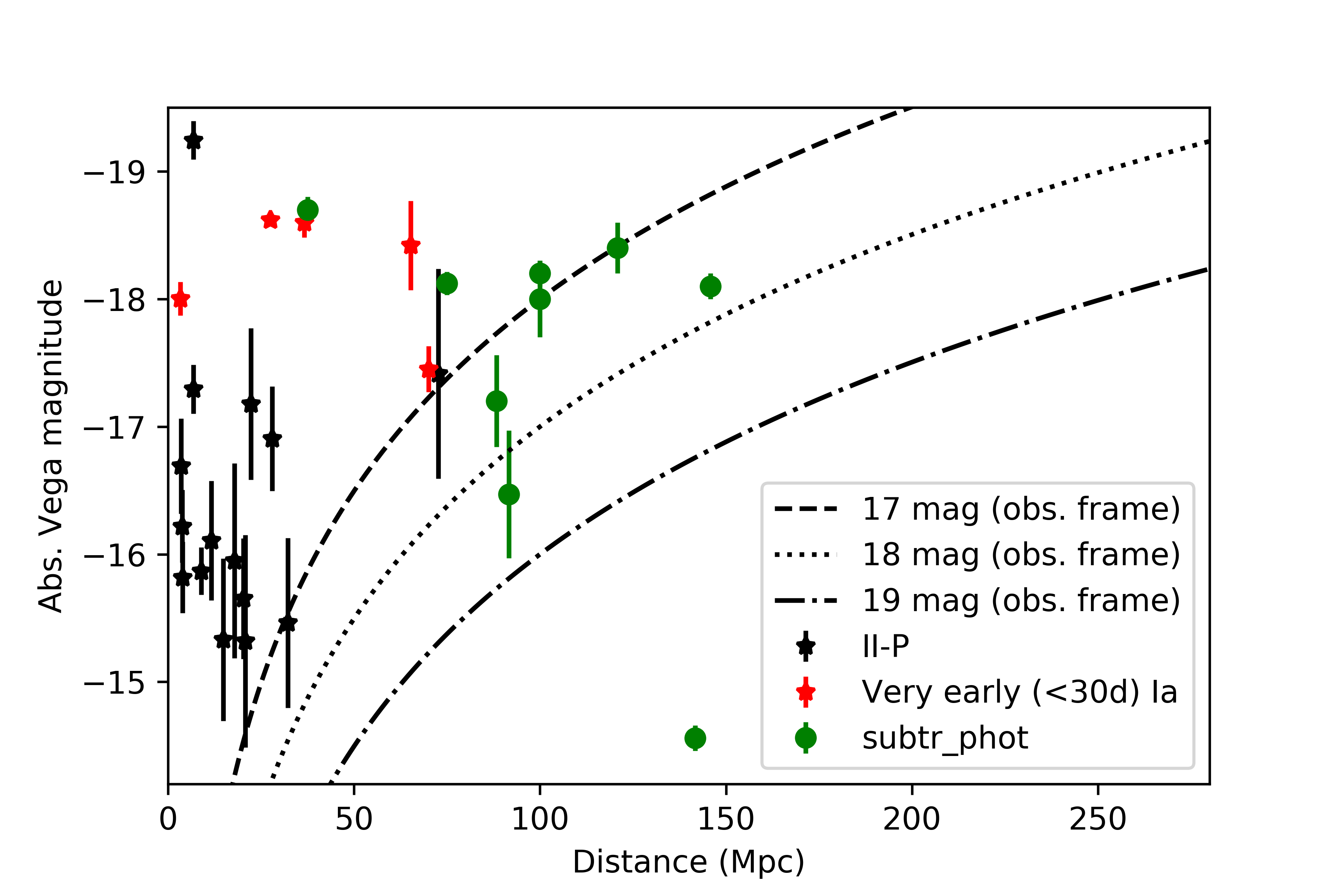}}
    }
\caption{Absolute magnitudes of the nine SNe discovered in our sample (green symbols).  The photometry corresponds to zero host galaxy extinction for all the events and therefore their absolute magnitudes should be considered as lower limits.  Also plotted are early (i.e.,$<3$\,month) photometry from \citet{szalai19} of nearby ($< 50$\,Mpc) SNe~IIP and some SNe~Ia within 30 days of explosion (black and red symbols, respectively). Dashed, dotted, and dotted-dashed lines represent apparent brightness levels of 17, 18, and 19\,mag, respectively.}
\label{fig_detection_mags}
\end{figure}

Table~\ref{tab_detections} lists, and Figures~\ref{fig_detection_images} and \ref{fig_detection_mags} show, all \detections\ detections from our sample.  We also indicate any corresponding ground-based discoveries, which help us constrain the explosion dates.  SN 2013if was discovered by the Supernova UNmasked By Infra-Red Detection (SUNBIRD) project and had an associated near-IR light curve \citep{kool18}.  The best-fit light-curve template shows SN 2013if to be an SN~IIP with almost no extinction, which is surprising given its proximity to the nucleus. \citet{kankare21} reported the discovery of PSN J10275082-4354034 near the nucleus of NGC 3256.  The SN was also detected in serendipitous {\it Hubble Space Telescope}~high-resolution imaging.  The subsequent light curve was determined to be most consistent with an SN~IIn, although an SN~IIP could not be ruled out.  Again, only a small amount of extinction ($A_V\lesssim0.3$\,mag) is present.  PS1-14xw was first reported by \citet{benitez14} and classified as an SN~Ia, while SN 2013dc was first reported by \citet{block13}.  We find no published analysis of either PS1-14xw or SN 2013dc.

Figure \ref{fig_mir_lc} plots the MIR light curves along with previously published light curves from \citet{szalai19}.  The MIR evolution of SNe~Ia and SNe~IIP overlap substantially, especially at early times, but they do tend to diverge a bit more at later times.  Of the light curves that extend out to later epochs, the evolution is most consistent with SNe~IIP.  Of course, PS1-14xw was originally classified as an SN~Ia.  For the SNe for which we only have a single epoch, the brightness is consistent with both SNe~Ia and SNe~IIP around peak with very little to no extinction.  The detection in Arp148 stands out for being relatively faint, which could indicate significant extinction or perhaps a poor constraint on the explosion time.  It is difficult to draw any conclusions on the extinction from just a single filter.

Even with the forward modeling, all of our SN detections are outside ($> 3$\arcsec) of the galactic nucleus.  (SN 2013if is a nuclear event included in our tally for rates purposes, but we do not have a convincing detection in Figure \ref{fig_detection_images}.)  The survey has limited follow-up observations to confirm the nature and classification of each transient source, but the observed magnitudes are consistent with those of other SNe (Fig.~\ref{fig_detection_mags}), bright enough to rule out most other types of transients (see Fig. 4 of \citealt{kasliwal17}).  For the purposes of this analysis, we assume that each detection (except PS1-14xw) is a CCSN.

\subsection{Survey Sensitivity}
\label{sec:sensitivity}

Before we can interpret the statistics of our detections, it is important to have an understanding of our survey's sensitivity.  We first consider our search sensitivity from an empirical perspective.  Figure~\ref{fig_detection_mags} plots the photometry of SNe detected in our survey (green points), as well as the relatively early ($< 3$\,month) magnitudes of the nearby ($< 50$\,Mpc) CCSNe from Figure~\ref{fig_lumfunction} (black points) and some young SNe~Ia from \citet{szalai19} (red points).  Overplotted are lines of constant apparent magnitude.  The early-time magnitude distribution of nearby CCSNe is $-16 \lesssim M_{3.6\,\mu m} \lesssim -18$, while known SNe~Ia are a couple of magnitudes brighter.  We assume that the nearby distribution of magnitudes is relatively complete.  If our survey were also to be considered complete and the magnitude distribution doesn't change as a function of redshift, then to first order we should detect a similar spread in our SN magnitude distribution out to 200\,Mpc.

\begin{figure}
\centering
\vbox{
    \raisebox{0.0in}{\includegraphics[width=3.5in]{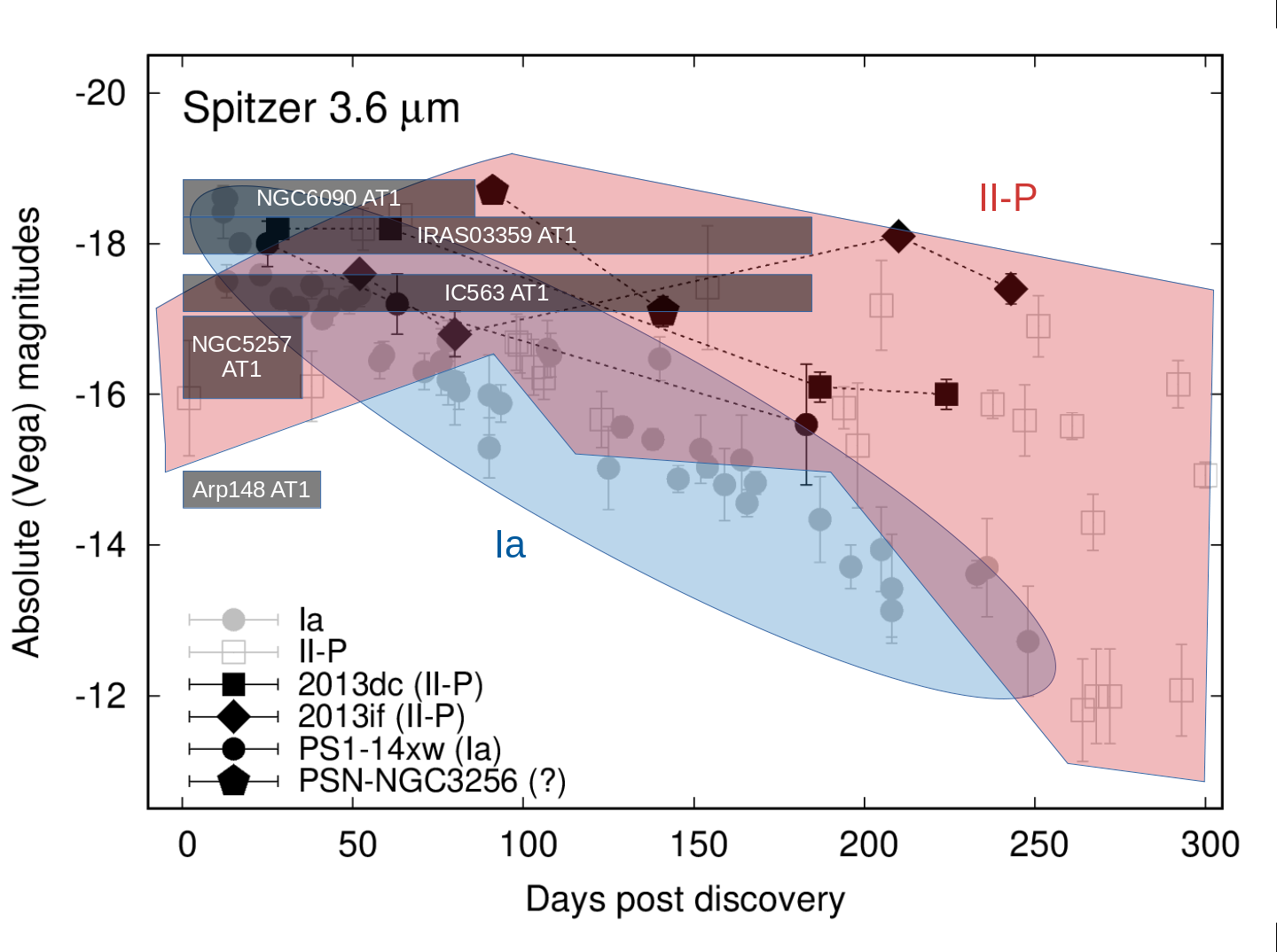}}
    }
\caption{The 3.6 $\mu$m light curves of the four detected known transients in our Survey (with black symbols), together with previously published light-curves \citep[grey symbols and shaded  colored regions, adopted from][]{szalai19}. Dark-grey transparent rectangles mark the single epoch of photometry for the other five detected sources within the possible periods (taking into account the dates of last non-detections). Widths and heights of the rectangles denote the uncertainties of epochs and absolute magnitudes of the five previously unknown transients, respectively.}
\label{fig_mir_lc}
\end{figure}

The actual distribution of our SN photometry in Figure~\ref{fig_detection_mags}, however, does not reflect the more nearby distribution.  Most of the SNe we detect are relatively bright.  In fact, as noted in Figure \ref{fig_mir_lc}, an absolute magnitude $M_{\rm 3.6\,\micron} \approx -18$ may be interpreted as more consistent with the peak magnitudes of SNe~Ia rather than those of SNe~IIP.  Although our sample is limited in size, we conclude that we are Malmquist biased and generally do not detect SNe dimmer than $17.5 \pm 0.5$\,mag.  This empirical limit is compared to the results of our artificial-star tests in Figure \ref{fig_cutoffmag}.  While in most cases the empirical threshold is brighter than the artificial-star tests achieved, there are a number of cases where the artificial-star tests are brighter than the empirical limit, particularly in the innermost regions of the galaxies.  As noted in Section~\ref{sec_algorithm}, there are a number of caveats in the sensitivity of the artificial-star tests to the fainter stars.  In general, we therefore define our sensitivity limit for a given radial bin to be the brighter of these two limits.

\subsection{Possible Causes for Decreased Sensitivity}

This survey is not particularly sensitive to SNe in either the nuclear regions of galaxies or galaxies at $>150$\,Mpc.  \citet{cresci07} offer some explanations for their {\it HST}/NICMOS survey that could apply to our survey.  For example, the FIR flux used to calculate the expected CCSNr is dominated by obscured AGNs and not by star formation. Estimates of the relative AGN power, however, have been made using a variety of MIR tracers in hundreds of local IR galaxies and, in most cases, do not suggest significant AGN contributions \citep[e.g.,][]{diaz-santos17}.

Another possibility is that the SNe may be more obscured by dust than originally expected, or underluminous SNe \citep[e.g.,][]{pastorello04} form a significant fraction of all CCSNe. In the latter case, the SNe would stay above our detection limit for a shorter time. In either case, the Gaussian distribution of magnitudes would have a tail skewed toward dimmer objects and our overall sensitivity would decrease, resulting in a decrease in our expected CCSNr.

\begin{figure}
\centering
\includegraphics[width=3.3in]{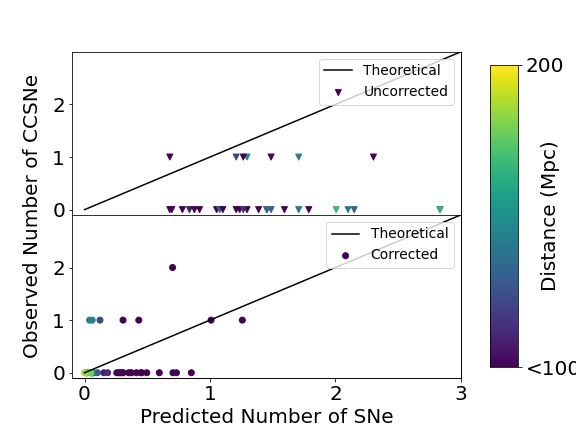}\\
\includegraphics[width=3.3in]{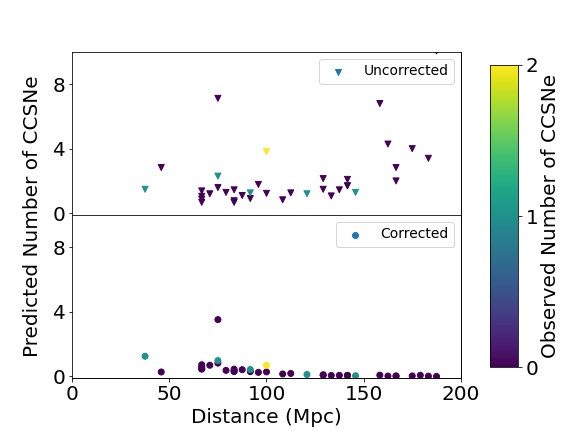}
\caption{Relationship between predicted CCSN rate, actual CCSN rate, and distance.  {\it (Top:)} The number of actual detections compared to the predicted number of CCSNe, where the colour bar corresponds to the distance (Mpc), for both the uncorrected and corrected rates. {\it (Bottom:)} The predicted number of SNe as a function of distance, where the colour bar corresponds to the number of actual detections. Both plots show values for both the uncorrected and corrected rates (as described in the text).}
\label{fig_expected_observed} 
\end{figure}

The most likely explanation, however, is a combination of worse than expected subtraction residuals combined with relatively poor resolution, despite the decreased extinction afforded by observations at 3.6\,\micron.  Note that the mosaic images have a pixel scale of $\sim 0.6''$ that corresponds to a projected distance of $\sim 290$\,pc at 100\,Mpc (and the real resolution is a factor of 2--3 worse).  Our poor residual at $<3$\arcsec\ corresponds to $\sim$1.5 kpc at these distances.  If most starburst activity is concentrated within the galaxy nucleus, then our survey not being sensitive to a majority of these events.  We correct for the limitations in our survey sensitivity below.

\begin{table}
\centering
\caption{IR Integrated Luminosities (5--24 \micron)}
\begin{tabular}{lrrrr}
Galaxy &  1\arcsec &  3\arcsec &  5\arcsec &  10\arcsec \\
Name & \multicolumn{4}{c}{\lsolar~(Fraction of Total Luminosity)}\\
\hline
NGC034      &    9.7 (0.1) &  10.4 (0.47) &  10.5 (0.62) &  10.7 (1.0) \\
NGC232      &   9.3 (0.05) &   9.9 (0.22) &  10.2 (0.44) &  10.6 (1.0) \\
MCG+12-02   &   9.2 (0.04) &  10.0 (0.24) &  10.3 (0.47) &  10.7 (1.0) \\
IC1623      &   8.9 (0.02) &   9.8 (0.15) &  10.1 (0.32) &  10.6 (1.0) \\
UGC2369     &   9.0 (0.02) &   9.6 (0.11) &  10.0 (0.25) &  10.6 (1.0) \\
IRAS03359   &   9.7 (0.07) &  10.3 (0.32) &  10.5 (0.46) &  10.8 (1.0) \\
MCG-03-12   &   9.4 (0.05) &  10.0 (0.27) &  10.3 (0.45) &  10.6 (1.0) \\
NGC1572     &   9.4 (0.09) &   10.1 (0.4) &  10.2 (0.52) &  10.5 (1.0) \\
NGC1614     &   9.5 (0.05) &  10.4 (0.38) &  10.6 (0.59) &  10.8 (1.0) \\
NGC2623     &   9.6 (0.09) &  10.2 (0.34) &   10.3 (0.5) &  10.6 (1.0) \\
UGC4881     &   9.7 (0.06) &  10.3 (0.26) &  10.5 (0.42) &  10.9 (1.0) \\
UGC5101     &  10.2 (0.11) &  10.8 (0.44) &  10.9 (0.56) &  11.1 (1.0) \\
MCG+08-18   &   8.2 (0.02) &   9.1 (0.16) &   9.3 (0.31) &   9.8 (1.0) \\
IC563       &   8.8 (0.03) &    9.6 (0.2) &   9.8 (0.36) &  10.3 (1.0) \\
NGC3110     &   9.1 (0.05) &    9.9 (0.3) &  10.1 (0.46) &  10.4 (1.0) \\
NGC3256     &   9.3 (0.04) &  10.1 (0.23) &  10.4 (0.43) &  10.7 (1.0) \\
IRAS10565   &   10.2 (0.1) &  10.8 (0.43) &  11.0 (0.54) &  11.2 (1.0) \\
Arp148      &   9.7 (0.08) &  10.3 (0.34) &  10.5 (0.49) &  10.8 (1.0) \\
MCG+00-29   &   9.6 (0.09) &  10.3 (0.42) &  10.4 (0.55) &  10.6 (1.0) \\
IC2810      &   9.6 (0.07) &   10.2 (0.3) &  10.4 (0.45) &  10.8 (1.0) \\
NGC3690     &   9.3 (0.04) &  10.1 (0.26) &  10.4 (0.45) &  10.7 (1.0) \\
ESO507-G070 &   9.4 (0.07) &  10.1 (0.32) &  10.3 (0.47) &  10.6 (1.0) \\
UGC8335     &   8.9 (0.12) &   9.6 (0.54) &   9.7 (0.78) &   9.8 (1.0) \\
UGC8387     &   9.7 (0.07) &  10.4 (0.35) &  10.6 (0.52) &  10.8 (1.0) \\
NGC5256     &   9.0 (0.02) &   9.8 (0.13) &  10.2 (0.35) &  10.7 (1.0) \\
NGC5257     &   9.0 (0.04) &   9.6 (0.17) &   9.9 (0.32) &  10.4 (1.0) \\
Mk273       &  10.4 (0.09) &  11.0 (0.35) &   11.1 (0.5) &  11.4 (1.0) \\
NGC5331     &   9.6 (0.06) &  10.3 (0.32) &  10.5 (0.47) &  10.8 (1.0) \\
UGC8782     &   9.3 (0.08) &   9.9 (0.35) &  10.1 (0.48) &  10.4 (1.0) \\
Arp302      &   9.7 (0.06) &   10.4 (0.3) &  10.6 (0.48) &  10.9 (1.0) \\
Mk848       &   10.0 (0.1) &  10.7 (0.41) &  10.8 (0.54) &  11.1 (1.0) \\
Arp220      &   9.8 (0.15) &  10.4 (0.69) &  10.5 (0.83) &  10.6 (1.0) \\
NGC6090     &   9.1 (0.02) &   9.9 (0.14) &   10.2 (0.3) &  10.7 (1.0) \\
NGC6240     &  10.0 (0.11) &  10.7 (0.47) &   10.8 (0.6) &  11.0 (1.0) \\
IRAS17208   &  10.4 (0.08) &  11.0 (0.35) &  11.2 (0.49) &  11.5 (1.0) \\
IC4687      &   9.3 (0.05) &  10.1 (0.34) &  10.4 (0.55) &  10.6 (1.0) \\
IRAS18293   &   9.7 (0.06) &  10.5 (0.32) &   10.7 (0.5) &  11.0 (1.0) \\
NGC6926     &   8.8 (0.04) &   9.5 (0.18) &   9.7 (0.32) &  10.2 (1.0) \\
NGC7130     &   9.4 (0.08) &  10.1 (0.34) &  10.2 (0.49) &  10.5 (1.0) \\
IRAS23128   &  10.1 (0.09) &  10.8 (0.38) &  10.9 (0.54) &  11.2 (1.0) \\
\end{tabular}

\label{tab_spitzerflux}
\end{table}

\subsection{Observed vs. Expected}

To properly compare the observed number of SNe detected to the predicted value, we first have to correct for the number of expected SNe to account for our decreased sensitivity.  To properly correct, we must calculate both the fraction of SNe we are sensitive to at any given location in a galaxy, $f_{\rm SNe}$, and the fraction of light within that radius, $f_{\rm light}$.

We first define our detection threshold in Section \ref{sec:sensitivity}.  We then assume the distribution of SN magnitudes to be consistent with observations in Figure \ref{fig_lumfunction}, which we take to be a Gaussian centred function at $M_{\rm 3.6\,\mu m} = -17$\,mag and a standard deviation of $\sigma_{\rm SN~IIP} = 0.7$.  No extinction is applied.  This distribution should be considered an upper limit as the actual function may have a lower-luminosity tail or be shifted to a somewhat less-luminous centre in the case of high extinction.  For each galaxy, we calculate the respective apparent magnitude distribution for the given distance and the corresponding fraction of SNe, $f_{\rm SNe}$, in the magnitude distribution that is bright enough to be detected based on the detection threshold.  This fraction is calculated separately for each radial bin.

We next derive the fraction of light within each radial bin, $f_{\rm light}$, from the integrated MIR lumninosity, which we measure using aperture photometry performed on archival IRAC and MIPS data (when available).  Although the SFR and the CCSNr are tied to the FIR luminosity, we assume the MIR to be a useful and convenient proxy, at least to first order. While most (U)LIRG flux comes out at longer wavelengths (i.e., $\sim 70$\,\micron), the shape of a (U)LIRG SED is roughly constant \citep{wright14}.  Table \ref{tab_spitzerflux} lists our derived MIR luminosities and fractional luminosity for each aperture.

The predicted CCSNr is finally corrected by multiplying by both the fraction of SNe we are sensitive to, $f_{\rm SNe}$, and the fraction of light within that radial bin, $f_{\rm light}$.  Table \ref{tab1} lists the corrected CCSNr predictions, and Figure~\ref{fig_expected_observed} compares the predicted and observed number of SNe detected for each galaxy. Figure~\ref{fig_expected_observed}(a) compares the number of observed CCSNe to the number of predicted CCSNe.  Overplotted is a line corresponding to a 1:1 ratio, which is what we would expect in an ideal scenario. The two plots correspond to the uncorrected and corrected number of predicted events, respectively.   Figure~\ref{fig_expected_observed}(b) plots the predicted number of CCSNe as a function of distance.  Again, the upper and lower panels correspond to the uncorrected and corrected number of predicted events, respectively.


There are several qualitative takeaways.  First, all of our detections occur at $<150$\,Mpc.  Although our calculations suggest that we should be sensitive to SNe out to 200\,Mpc, other challenges arise at these distances.  For example, galaxies become less resolved and more compact, so the subtraction residuals have a larger impact.  These residuals correspond to larger projected distances in, and therefore fraction of, the host galaxy.  While there is no sharp cutoff, the more distant galaxies in our sample are more luminous, but also more compact and dustier.  

Second, in almost every case the adjusted expected CCSNr ends up falling to $<1$, which makes a comparison to the expected trend (black line) in Figure~\ref{fig_expected_observed}(a) difficult.  Despite small-number statistics, we still see a slight correlation between observed SNe versus predicted SNe that, although weak, does provide a useful self-consistency check.  The one exception is Arp 220, which stands out from the other galaxies at $\sim75$\,Mpc for its large number of predicted SNe.  Despite this expectation, we don't detect any SNe, possibly due to the fact that most of the light is concentrated in the inner $<3$\arcsec.  Furthermore, like other IR-bright galaxies, Arp 220 may suffer from relatively high local extinction.  

Taken all together, these adjustments limit a significant fraction of the phase space in which we can search for and detect an SN.  The large number of galaxies observed, however, compensates for these inefficiencies and provides adequate statistics.  In total, we expect 96 and \predicteddetections\ SNe before and after corrections are made (respectively), where the error bars are dominated by uncertainty assigned above to our detection threshold (i.e., $m_{\rm 3.6\,\mu m}=17.5\pm0.5$\,mag).  We discover \detections\ SNe (Table~\ref{tab_detections}).

\subsection{Statistical Implications}

After all corrections are taken into account, our observations yield a statistically significant sample that will enable us to differentiate between (1) undetected SNe and (2) counting statistics.  Assuming the counting statistics are determined by the SN sample (i.e., $\sigma=\sqrt{N}$), then the minimum number of SNe we must expect to detect is {\it roughly} sixteen.  Any fewer expected SNe would be statistically insufficient.  For example, to identify a $3\sigma$ deficit in the CCSNr when only 9 SNe are expected, one would need to observe $<0$ SNe.

As noted above, we detect \detections\ SNe, whereas theory predicted \predicteddetections\ SNe.  We conclude that our observations are generally consistent with the predictions, especially when considering the counting statistics, which for 14 SNe yield $\sigma_{\rm counting}=\sqrt{14}=3.7$.  Of course, the number of predicted SNe is given {\it after} significant corrections in our sensitivity have been applied.

\section{Conclusion} \label{sec_conclusions}

We have presented a {\it Spitzer}/IRAC MIR survey for dust-extinguished SNe to resolve the discrepancy between theoretically predicted and optical/NIR observed core-collapse SN rates.  We searched 40 nearby ($< 200$\,Mpc) star-forming galaxies using {\it Spitzer}/IRAC.  The survey includes eight epochs of observations obtained in the years 2012--2014.

Our ability to detect new SNe was dominated by subtraction residuals near the nuclear regions ($<$3 \arcsec) of each galaxy caused by an asymmetric PRF that rotated throughout the year as the telescope circled the Sun.  To optimise our subtraction algorithm, we implemented a forward modeling technique that will be employed on the {\it NGRST}; \citep{Rubin2021}.  

While forward modeling improves our overall sensitivity, the number of discovered SNe is still substantially lower than the expected intrinsic number during the survey period.  We compensate for the number of predicted SNe by quantifying the inefficiencies, mostly arising from remaining residuals in the inner $<3$\arcsec, where a majority of star formation occurs.  The predicted rates are particularly impacted by the fact that most of the galaxies with the highest SN rates (2--11 SNe per year) are located at distances larger than 150\,Mpc, where essentially zero SNe can be expected to be discovered owing to the sensitivity limits of the search (Table \ref{tab1}).  After all corrections are applied, we expect to discover \predicteddetections\ SNe and actually find \detections\ SNe, which suggests that our observations are consistent with the theoretical expectations.  

While still hampered by high extinction, if most starburst activity is concentrated within the inner nucleus of the galaxy (for this experiment $<$1.5\,kpc at 100 Mpc), a survey with both higher resolution and longer-wavelength observations is necessary.  A stable, space-based PSF is needed, too.  Figure~\ref{fig_extinction} shows that the optimal wavelength is actually closer to 2.7\,\micron.  With the upcoming launch of {\it JWST}, with a PSF of $\sim 0.091$\arcsec\ in its F277W filter, such a survey to probe the nuclear regions of star-forming galaxies for transients will finally be possible.

\section*{Acknowledgements}

The authors acknowledge an anonymous referee for helpful comments that led to improvements in this paper. We also give tremendous thanks to the IRAC instrument support team at the {\it Spitzer} Science Center (SSC) for their continued assistance over the years. They are also grateful to Rick Arendt for his helpful comments. O.D.F. additionally thanks Peter Capak, Sean Carey, and Patrick Lowrance for useful technical discussions. This work is based on observations made with the {\it Spitzer Space Telescope} (PID 90031), which is operated by the Jet Propulsion Laboratory, California Institute of Technology, under a contract with NASA. Support for this work was provided by NASA through an award issued by JPL/Caltech, primarily supporting students C.C., G.L., and H.K. Additional support was provided to D.R. and H.K. by STScI's DDRF Grant D0001.82477 and the TSRC program.  T.S. is supported by the J\'anos Bolyai Research Scholarship of the Hungarian Academy of Sciences, by the New National Excellence Program (UNKP-20-5) of the Ministry of Technology and Innovation of Hungary, by the GINOP-2-3-2-15-2016-00033 project of the National Research, Development, and Innovation Office of Hungary (NKFIH) funded by the European Union, and by NKFIH/OTKA FK-134432 grant. A.V.F. acknowledges support from the TABASGO Foundation, the Christopher R. Redlich Fund, and the U.C. Berkeley Miller Institute for Basic Research in Science (in which he is a Senior Miller Fellow).

\bigskip
\noindent{\bf Data availability}\\

The data underlying this article will be shared on reasonable request to the corresponding author.\\

\bibliographystyle{mnras}
\bibliography{references}

\end{document}